\documentclass{icrc29}
\usepackage{graphicx,amssymb,amsmath,times}
\usepackage{psfrag}
\begin{document}
\title[Fluctuations in the EAS radio signal]
{Fluctuations in the EAS radio signal derived with improved Monte Carlo simulations based on CORSIKA}
\author[T. Huege et al.] {T. Huege$^{a,e}$, 
W.D.~Apel$^{a}$, 
F.~Badea$^{a}$,
L.~B\"ahren$^{b}$,
K.~Bekk$^{a}$, 
A.~Bercuci$^{c}$,
M.~Bertaina$^{d}$, 
\newauthor
P.L.~Biermann$^{e}$,
J.~Bl\"umer$^{a,f}$,
H.~Bozdog$^{a}$,
I.M.~Brancus$^{c}$,
S.~Buitink$^{g}$,
M.~Br\"uggemann$^{h}$,
\newauthor
P.~Buchholz$^{h}$,
H.~Butcher$^{b}$,
A.~Chiavassa$^{d}$,
K.~Daumiller$^{a}$, 
A.G.~de~Bruyn$^{b}$,
C.M.~de~Vos$^{b}$,
\newauthor
F.~Di~Pierro$^{d}$,
P.~Doll$^{a}$, 
R.~Engel$^{a}$,
H.~Falcke$^{b,e,g}$,
H.~Gemmeke$^{i}$, 
P.L.~Ghia$^{j}$,
R.~Glasstetter$^{k}$, 
\newauthor
C.~Grupen$^{h}$,
A.~Haungs$^{a}$, 
D.~Heck$^{a}$, 
J.R.~H\"orandel$^{f}$, 
A.~Horneffer$^{e,g}$,
K.-H.~Kampert$^{k}$,
\newauthor
G.W.~Kant$^{b}$, 
U.~Klein$^{l}$,
Y.~Kolotaev$^{h}$,
Y.~Koopman$^{b}$, 
O.~Kr\"omer$^{i}$, 
J.~Kuijpers$^{g}$,
S.~Lafebre$^{g}$,
\newauthor
G.~Maier$^{a}$,
H.J.~Mathes$^{a}$, 
H.J.~Mayer$^{a}$, 
J.~Milke$^{a}$, 
B.~Mitrica$^{c}$,
C.~Morello$^{j}$,
G.~Navarra$^{d}$,
\newauthor
S.~Nehls$^{a}$,
A.~Nigl$^{g}$,
R.~Obenland$^{a}$,
J.~Oehlschl\"ager$^{a}$, 
S.~Ostapchenko$^{a}$, 
S.~Over$^{h}$,
H.J.~Pepping$^{b}$,
\newauthor
M.~Petcu$^{c}$, 
J.~Petrovic$^{g}$,
T.~Pierog$^{a}$, 
S.~Plewnia$^{a}$,
H.~Rebel$^{a}$, 
A.~Risse$^{m}$, 
M.~Roth$^{f}$, 
H.~Schieler$^{a}$, 
\newauthor
G.~Schoonderbeek$^{b}$
O.~Sima$^{c}$, 
M.~St\"umpert$^{f}$, 
G.~Toma$^{c}$, 
G.C.~Trinchero$^{j}$,
H.~Ulrich$^{a}$,
\newauthor
S.~Valchierotti$^{d}$,
J.~van~Buren$^{a}$,
W.~van~Capellen$^{b}$,
W.~Walkowiak$^{h}$,
A.~Weindl$^{a}$,
S.~Wijnholds$^{b}$,
\newauthor
J.~Wochele$^{a}$, 
J.~Zabierowski$^{m}$,
J.A.~Zensus$^{e}$,
D.~Zimmermann$^{h}$\\
(a) Institut\ f\"ur Kernphysik, Forschungszentrum Karlsruhe,
76021~Karlsruhe, Germany\\
(b) ASTRON, 7990 AA Dwingeloo, The Netherlands \\
(c) National Institute of Physics and Nuclear Engineering,
7690~Bucharest, Romania\\
(d) Dipartimento di Fisica Generale dell'Universit{\`a},
10125 Torino, Italy\\
(e) Max-Planck-Institut f\"ur Radioastronomie,
53010 Bonn, Germany \\
(f) Institut f\"ur Experimentelle Kernphysik,
Universit\"at Karlsruhe, 76021 Karlsruhe, Germany\\
(g) Department of Astrophysics, Radboud University Nijmegen, 6525
ED Nijmegen, The Netherlands \\
(h) Fachbereich Physik, Universit\"at Siegen, 57068 Siegen, 
Germany \\
(i) Inst. Prozessdatenverarbeitung und Elektronik, 
Forschungszentrum Karlsruhe, 76021~Karlsruhe, Germany \\
(j) Istituto di Fisica dello Spazio Interplanetario, INAF,
10133 Torino, Italy \\
(k) Fachbereich Physik, Universit\"at Wuppertal, 42097
Wuppertal, Germany \\
(l) Radioastronomisches Institut der Universit\"at Bonn, 
53121 Bonn, Germany \\
(m) Soltan Institute for Nuclear Studies, 90950~Lodz, 
Poland\\
}
\presenter{Presenter: T. Huege (tim.huege@ik.fzk.de), \  
ger-huege-T-abs1-he14-oral}

\maketitle

\begin{abstract}
Cosmic ray air showers are known to emit pulsed radio emission which can be understood as coherent geosynchrotron radiation arising from the deflection of electron-positron pairs in the earth's magnetic field. Here, we present simulations carried out with an improved version of our Monte Carlo code for the calculation of geosynchrotron radiation. Replacing the formerly analytically parametrised longitudinal air shower development with CORSIKA-generated longitudinal profiles, we study the radio flux variations arising from inherent fluctuations between individual air showers. Additionally, we quantify the dependence of the radio emission on the nature of the primary particle by comparing the emission generated by proton- and iron-induced showers. This is only the first step in the incorporation of a more realistic air shower model into our Monte Carlo code. The inclusion of highly realistic CORSIKA-based particle energy, momentum and spatial distributions together with an analytical treatment of ionisation losses will soon allow simulations of the radio emission with unprecedented detail. This is especially important to assess the emission strengths at large radial distances, needed for event-to-event comparisons of the radio signals measured by LOPES in conjunction with KASCADE-Grande and for considerations regarding large arrays of radio antennas intended to measure the radio emission from ultra-high energy cosmic rays, as with LOFAR or in the Pierre Auger Observatory.
\end{abstract}

\section{Introduction}

As part of LOPES \cite{LOPES}, a digital radio interferometer working in conjunction with the KASCADE-Grande experiment, we have developed a model of the emission mechanism responsible for the radio pulses produced by extensive air showers. Describing the emission as coherent geosynchrotron radiation arising from the deflection of electron-positron pairs in the earth's magnetic field, we have performed analytical calculations \cite{HuegeFalcke2003} followed by detailed Monte Carlo simulations \cite{MonteCarlo}, which were then used to derive important dependences of the radio emission on specific air shower parameters \cite{HuegeFalcke2005}. For the first time, our simulations have employed a realistic model of the air shower producing the radio emission. Important air shower properties such as the longitudinal (``arrival time'') and lateral particle distributions in the shower pancake, the particle energy and track length distributions and the overall longitudinal evolution of the shower were taken into account via ``standard'' analytical parametrisations such as NKG functions (lateral distributions) and Greisen parametrisations (longitudinal evolution). While a number of important results have already been obtained with this model \cite{HuegeFalcke2005}, the next step in improving the simulations is to substitute the underlying air shower model by a more realistic one based on histogrammed distributions derived from CORSIKA \cite{CORSIKA} simulations.

Here, we present a first result derived with these improved simulations, namely the influence of the air showers' realistically simulated longitudinal profiles. These realistic CORSIKA-generated profiles allow us to evaluate shower-to-shower fluctuations in the radio emission as well as the influence of the nature of the primary particle on the radio signal.

\section{Discussion}

To assess the relative fluctuations in the radio signal between individual air showers, we have simulated 10 very weakly thinned proton induced air showers and 10 very weakly thinned iron induced air showers with a primary particle energy of $10^{16}$~eV and vertical incidence. The geomagnetic field was adopted the same as in \cite{HuegeFalcke2005}, namely 70$^{\circ}$ inclined with a strength of 0.5~Gauss, which approximately corresponds to the geomagnetic field configuration in Central Europe. CORSIKA was used with the GHEISHA2002 model \cite{GHEISHA} for low-energy hadronic interactions and the QGSJET01 model \cite{QGSJET} for high-energy hadronic interactions. The relatively low primary particle energy was chosen to minimise the computation time in this early stage of analysis. We will reconsider the effects for higher energies at a later time in conjunction with a much broader analysis of the improvements achieved with the adoption of the other CORSIKA-based particle distributions.

For the proton and iron showers each, the 10 simulations were averaged to obtain an ``average air shower'' with which one can then compare the individual showers. For this purpose, the individual showers were reviewed by eye and the cases with longitudinal profiles differing most from the average profile were selected. Simulations of the associated radio signal were then performed with our Monte Carlo code. Only the longitudinal profile of the shower development was taken from the CORSIKA simulations. All other air shower properties (energy distributions, spatial particle distributions, ...) were retained using the analytical distributions employed in \cite{HuegeFalcke2005}. This procedure allows a simple first estimate of the importance of the fluctuations in the shower development. For comparison, we also calculated the radio emission from a fully parametrised shower, retaining a Greisen parametrisation for the longitudinal development.

   \begin{figure}[!ht]
   \psfrag{Eomega0muVpmpMHz}[c][b]{$|E(\vec{R},2\pi\nu)|$~[$\mu$V~m$^{-1}$~MHz$^{-1}$]}   
   \centering
   \includegraphics[width=5.35cm,angle=270]{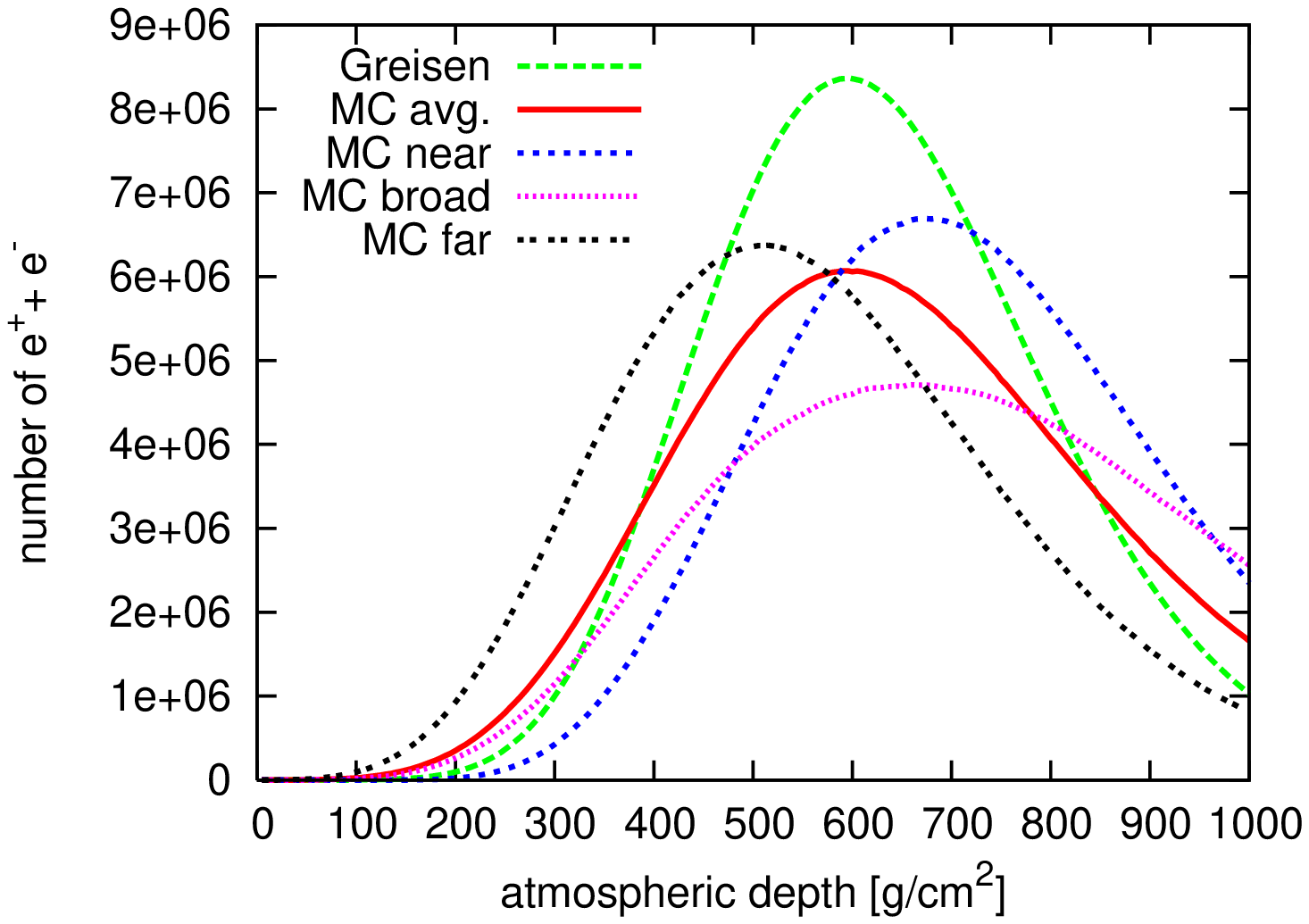}
   \includegraphics[width=5.35cm,angle=270]{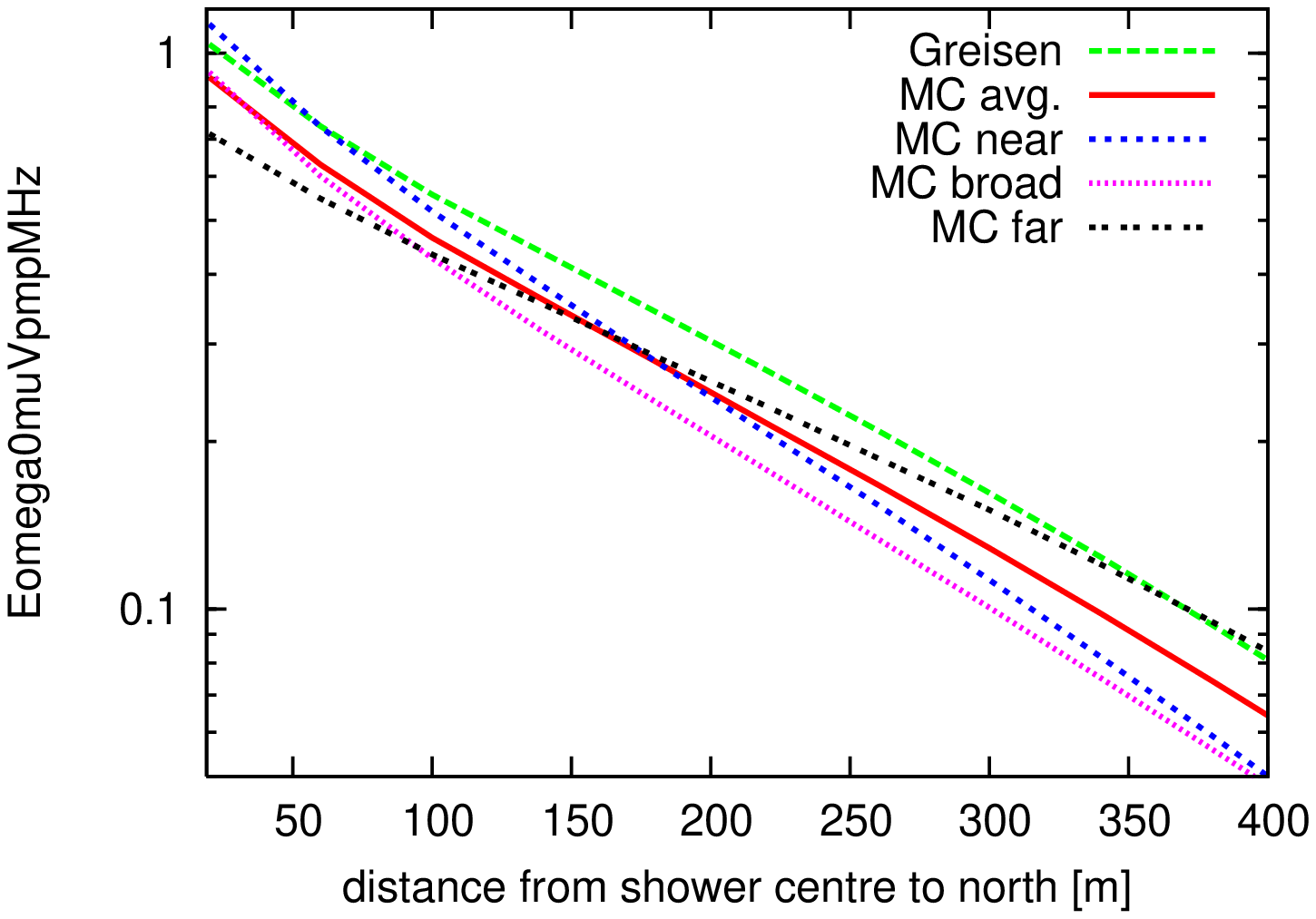}
   \caption[Shower to shower fluctuations]{
   \label{fig:profiles}
   Left: Longitudinal development profiles of vertical CORSIKA-simulated $10^{16}$~eV proton-induced air showers in comparison with an averaged profile derived from CORSIKA and a Greisen-parametrised profile. Right: Simulated radial dependence of the 10~MHz radio signal as calculated using the longitudinal profiles displayed on the left.
   }
   \end{figure}

The left panel of figure \ref{fig:profiles} shows a comparison of the longitudinal profiles taken into account for the analysis of the proton induced showers. In particular, the analytical Greisen parametrisation used so far in the code is very similar to the profile of the averaged proton shower. It does, however, overestimate the number of particles generated by $\sim$~15\%, as it describes a purely electromagnetic shower of the given energy without significant ``losses'' to a muonic component. The three selected individual shower profiles represent cases of a shower developing to its maximum far away from the ground, a shower developing to its maximum close to the ground, and a shower having a generally much broader maximum. The right panel of figure \ref{fig:profiles} shows the 10~MHz radio signal obtained from these longitudinal shower profiles. Due to the overestimation of the number of particles in the Greisen parametrisation, the emission is about $\sim$~15\% higher for the fully parametrised shower. Apart from this constant enhancement factor, the radial profile of the radio emission is very similar to that of the averaged proton shower, which indeed has a very similar longitudinal evolution. The individual showers, in comparison, show significant deviations from the radial profile generated by the average shower. While the overall emission level (directly related to the number of electrons and positrons) stays similar, the slope of the radial emission profile changes. The shower with the maximum close to the ground has a steeper slope, the one with the maximum far away from the ground has a flatter slope. This dependence of the radio emission on the depth of the shower maximum was already shown in \cite{HuegeFalcke2005}. The shower with the broader maximum has slightly lower radio emission over the radial distances considered here. This arises from the lower total number of electrons and positrons sampled in the atmosphere's $\sim$~1000~g~cm$^{-2}$ over the ground, after which the profile is ``truncated''. Overall, the deviations in the field strengths are of the order $\sim$~30\%.

   \begin{figure}[!ht]
   \psfrag{Eomega0muVpmpMHz}[c][t]{$|E(\vec{R},2\pi\nu)|$~[$\mu$V~m$^{-1}$~MHz$^{-1}$]}   
   \centering
   \includegraphics[width=0.95\textwidth]{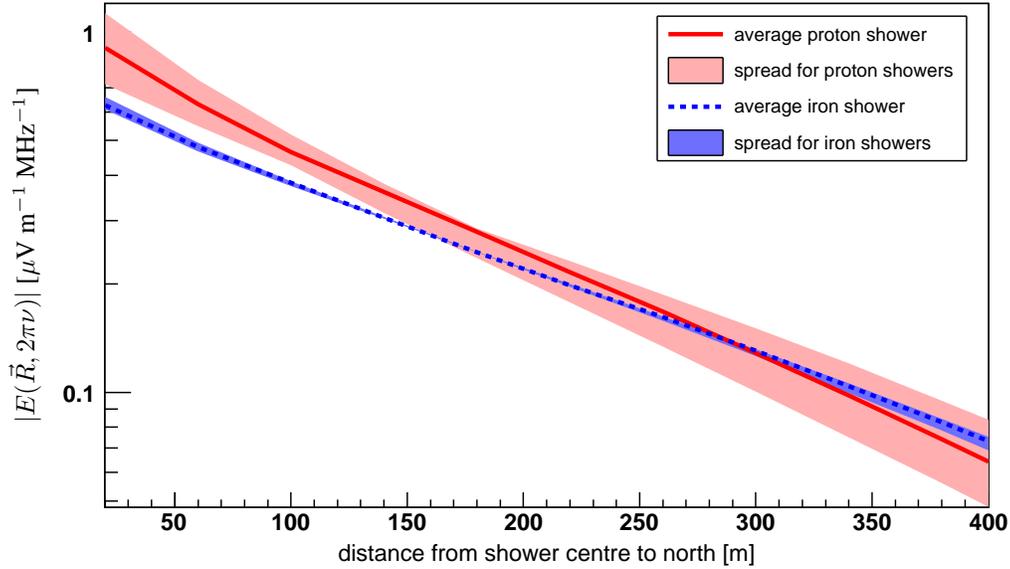}
   \caption[Chemical composition dependence]{
   \label{fig:overall}
   Radial dependence of the 10~MHz radio emission from vertical $10^{16}$~eV air showers induced by protons and iron nuclei, respectively, comparing the emission from an averaged shower with that from individual showers.
   }
   \end{figure}

For the iron-induced showers, the fluctuations in the longitudinal shower development are much smaller, as expected from the simplified ``superposition model'' interpreting an iron-induced shower as a superposition (and thus averaging) of 56 nucleon-induced showers. Correspondingly, the deviations of the radio signal between the ``average shower'' and the individual showers are much smaller. This is shown in figure \ref{fig:overall} in comparison with the fluctuations for the proton-induced showers. As the iron-induced showers grow to their maxima at larger distances from the observer than the proton induced showers, the radial profiles of their radio emission are somewhat flatter. Additionally, the emission amplitudes are slightly lower. This is due to the fact that for iron-induced showers, the fraction of energy going into the muonic shower component is higher (and thus the fraction going into the electromagnetic component lower) than in proton-induced showers.

\section{Conclusions}

Using CORSIKA-generated air shower development profiles, we have analysed the influence of shower-to-shower fluctuations on the associated radio signal. While vertical $10^{16}$~eV showers induced by proton primaries show fluctuations of the order 30\%, the fluctuations for iron-induced showers with the same parameters are negligible. Two characteristics of the shower profiles have a direct impact on the radio emission: the distance of the shower maximum from the observer on the ground, directly influencing the slope of the radial dependence, and the total number of electrons and positrons integrated over the shower evolution, directly determining the absolute strength of the emission. Differences in the radio emission of individual showers and also differences in the emission from showers induced by different primaries can be related directly to these two effects.

The analysis we performed constitutes a first step towards composition studies of cosmic rays with radio techniques. However, it was carried out with only limited statistics at relatively low energies, and only for vertical air showers. A more complete analysis at higher energies and various zenith angles will be performed soon in conjunction with a detailed analysis of the effects introduced by substituting other analytical parametrisations such as those describing energy and pitch-angle distributions by more realistic CORSIKA-derived data. It is, however, clear that the influence of differences in the longitudinal shower profiles on the radio emission will still mainly be related to the spatial distance of the shower maximum to the ground and the total number of electrons and positrons integrated over the shower evolution.

\end{document}